\documentclass[fleqn,usenatbib]{mnras}

\usepackage{mathptmx}

\usepackage[T1]{fontenc}
\usepackage{ae,aecompl}

\usepackage{graphicx}	
\usepackage{amsmath}	
\usepackage{amssymb}	
\usepackage{cases}  

\title[Flaring SPI in AU Mic]{Searching for flaring star-planet interactions in AU Mic TESS observations}

\author[E. Ilin et al.]{
E. Ilin$^{1,2,3}$,\thanks{E-mail: eilin@aip.de}
K. Poppenhaeger$^{1,2}$,
\\
$^{1}$Leibniz-Institute for Astrophysics Potsdam (AIP), An der Sternwarte 16, 14482 Potsdam, Germany\\
$^{2}$Institute for Physics and Astronomy, University of Potsdam, Karl-Liebknecht-Str. 24/25, 14476 Potsdam, Germany\\
$^{3}$Department of Astrophysics, American Museum of Natural History, 200 Central Park West, Manhattan, NY, USA\\
}

\date{Accepted XXX. Received YYY; in original form ZZZ}

\pubyear{2021}

\hypersetup{draft}
\begin{document}
\label{firstpage}
\pagerange{\pageref{firstpage}--\pageref{lastpage}}
\maketitle

\begin{abstract}
Planets that closely orbit magnetically active stars are thought to be able to interact with their magnetic fields in a way that modulates stellar activity. This modulation in phase with the planetary orbit, such as enhanced X-ray activity, chromospheric spots, radio emission, or flares, is considered the clearest sign of magnetic star-planet interaction (SPI). However, the magnitude of this interaction is poorly constrained, and the intermittent nature of the interaction is a challenge for observers. AU Mic is an early M dwarf, and the most actively flaring planet host detected to date. Its innermost companion, AU Mic b, is a promising target for magnetic SPI observations. We used optical light curves of AU Mic obtained by the Transiting Exoplanet Survey Satellite to search for signs of flaring SPI with AU Mic b using a customized Anderson-Darling test. In the about $50$ days of observations, the flare distributions with orbital, rotational, and synodic periods were generally consistent with intrinsic stellar flaring. We found the strongest deviation ($p=0.07,\;n=71$) from intrinsic flaring with the orbital period of AU Mic b, in the high energy half of our sample ($ED>1$\,s). If it reflects the true SPI signal from AU Mic b, extending the observing time by a factor of $2-3$ will yield a $>3\sigma$ detection. Continued monitoring of AU Mic may therefore reveal flaring SPI with orbital phase, while rotational modulation will smear out due to the star's strong differential rotation.
\end{abstract}


%

\section{Introduction}

Flares are electromagnetic explosions in the stellar corona that are driven by the dynamics of the surface magnetic fields~\citep{benz2010}. They can be triggered intrinsically by the star in isolation, but also by magnetic star-planet interaction (SPI). In SPI flares, the event is induced by the re-connection of stellar and planetary magnetic field lines~\citep{saur2013magnetic,lanza2018close-by,fischer2019}. For this to occur, the planet must revolve around the star in close orbit, so that it moves within the stellar Alfv\'en zone for at least a fraction of its orbit. When the planet is within the Alfv\'en zone, the magnetic field and the energy transported along the field lines as particles or waves can fall back on to the star, whereas outside of it it would be carried away by the stellar wind pressure. 

Several studies of individual systems with close-in and eccentric Hot Jupiters report excess flaring in phase with the planet or close to periastron~\citep{shkolnik2005hot,pillitteri2011,maggio2015}, but in light of many non-detections of excess flares in similar systems~\citep{figueira2016, fischer2019}, flaring SPI remains an elusive phenomenon. One interpretation is that the interaction is so weak and intermittent that the system has to be observed for many orbits of the interacting planet before SPI flares become measurable against the background of intrinsic, stochastic flares~\citep{shkolnik2008nature,lanza2009,saur2013magnetic,strugarek2015}.

Recently, AU Mic, an M0-M1~\citep{pecaut2013,gaidos2014} pre-main sequence dwarf star, and a member of the $\beta$ Pic young moving group, which is $16-29$ Myr old~\citep{malo2014,binks2014,mamajek2014,bell2015,binks2016,shkolnik2017,miretroig2020}, was discovered to be a favorable candidate for magnetic star-planet interactions~\citep{kavanagh2021}. The system hosts two planets in close orbits near a 2:1 mean-motion resonance, AU Mic b and AU Mic c~\citep{plavchan2020,martioli2021new}, with masses $M_b=11.7 \pm 5.0 M_\oplus$  and $M_c = 22.2 \pm 6.7 M_\oplus$~\citep{zicher2022one}. Several papers reported system parameters of AU Mic based on photometry from the Transiting Exoplanet Survey Satellite~\citep[TESS,][]{ricker2014}, the CHEOPS mission~\citep{benz2021cheops}, and HARPS~\citep{mayor2003setting} spectroscopy~\citep{plavchan2020, cale2021diving, gilbert2021flares, zicher2022one}, which are mutually compatible within uncertainties. In this work, we use the determinations by \citet{gilbert2021flares}, who based their analysis on the complete TESS photometry.

AU Mic rotates at a $4.85\pm0.03$ d period~\citep{gilbert2021flares}, showing strong solar-like differential rotation, a strong, mostly poloidal large-scale magnetic field of $>400$ G, and several activity indicators that vary in phase with the star's latitude-dependent rotation periods~\citep{klein2021}. Optical light curves obtained by TESS confirm previous observations~\citep{katsova1999, robinson2001, redfield2002} that AU Mic is actively flaring~\citep{martioli2021new}.

\mbox{AU Mic b} is a Neptune-sized planet (\mbox{$R_p = 4.19\pm0.24R_\oplus$},~\citealt{gilbert2021flares}) that was first discovered using TESS photometry~\citep{plavchan2020} with an orbital period of \mbox{$P_{orb}=8.4630004\pm0.000006$ d}~\citep{gilbert2021flares}. The system shows strong transit timing variations, but the mean orbital period $P_{mean} = 8.4631427 \pm 0.0000005$ d is similar enough to the instantaneous $P_{orb}$~\citep{szabo2022transit} for the purpose of this study. 
\citet{kavanagh2021} predict magnetic SPI with AU Mic b to be observable in the radio regime, a signal indicative of a phenomenon similar to the planet-moon interaction observed within the Jupiter-Io system~\citep{saur2013magnetic}. \citet{kavanagh2021} found that, based on reconstructions of AU Mic's magnetic field and mass loss rate, AU Mic b could reside in the sub-Alfv\'enic regime. If this is true, flaring SPI is also possible~\citep{lanza2018close-by}; a recent study even claimed to have visually identified hints of such an interaction in TESS data~\citep{colombo2022short}. In fact, AU Mic is the most actively flaring star among all currently known exoplanet hosts (Ilin et al. in prep.), implying that the footpoint of interaction may frequently pass regions where high magnetic tension has built up, and therefore trigger flares more easily than in other star-planet systems. 

Two further reasons favor AU Mic in the search for flaring SPI: First, AU Mic b's orbit is not synchronized with the stellar rotation, so that confusion with rotational modulation of intrinsic flaring activity is unlikely. Second, the system is observed equator-on. The footpoint of interaction would therefore move in and out view of the observer regardless of its latitude on the star.

When observational biases are taken into account, the absence of flaring SPI signal can be informative, too. It favors models with strong stellar winds that produce a small Alfv\'en zone~\citep{kavanagh2021}; weak planetary magnetic fields, and steady energy transfer channels between planet and star rather than eruptive, flaring ones~(see Section~\ref{sec:discussion}).  

In this work, we searched the TESS light curves of AU Mic for statistically reliable signs of flaring SPI. We present our light curve de-trending and flare finding method in Section \ref{sec:detrendfind}, and the resulting flare catalog in Section \ref{sec:flarecatalog}. In Section \ref{sec:phases}, we show how we can combine flare samples from general time series measurements into a homogeneous data set, and apply this method to test whether flaring SPI signal is present in the TESS observations. The main result is shown in Table~\ref{tab:pvals}. We discuss our results and present our conclusions in Sections~\ref{sec:discussion} and \ref{sec:conclusions}.
\section{Observations}
\subsection{TESS photometry}
The Transiting Exoplanet Survey Satellite~(TESS,~\citealt{ricker2014}) is an all-sky mission that began operations in 2018, and completed its first full sky scan in April 2020. It is still observing at the time of writing, collecting nearly continuous photometric times series in the 600-1000 nm band for $\sim 27$ d in each observing Sector. Light curves of $232\,763$ stars observed in 2 min cadence in the first two years of operations~(Sectors 1 to 26) are currently available on the Mikulski Archive for Space Telescopes\footnote{as of Mar 16, 2022}. From Sector 27 on, 1000 targets were observed at even higher 20 s cadence each Sector. AU Mic was observed in Sector 1 at a 2 min cadence, and in Sector 27 at a 20 s cadence. The light curves reveal strong rotational variability from starspots, vigorous flaring, and transits of two planets, AU Mic b and c~\citep{plavchan2020,martioli2021new}.

\subsection{Light curve de-trending and flare finding}
\label{sec:detrendfind}
Stellar light curves are time series of flux measurements that vary due to both astrophysical and instrumental effects. Flares are only one of many phenomena like spot variability, transits, eclipses, bursts and dips that can be detected in the data. Accurate automated flare detection algorithms are still challenging to design~\citep{vida2021}, not least due to this intrinsically heterogeneous morphology of light curves. We applied an iterative three-step algorithm to remove all but the variations caused by flares, and derived a realistic noise estimate (Section~\ref{sec:detrend}). We then used a $\sigma$-clipping procedure to mark flare candidates and calculate their properties~(Section~\ref{sec:flarefind}). The resulting flare catalog is presented in Section~\ref{sec:flarecatalog}.

\subsubsection{De-trending}
\label{sec:detrend}
Typical flare times range from a few minutes to a few hours, and rarely exceed one day in duration. Most stellar variability occurs on longer time scales, except for ultrafast rotational variability in some stars~\citep{ilin2021giant}. The light curve de-trending presented here was implemented with a broad spectrum of Kepler~\citep{borucki2010} and TESS light curve types in mind. Our method consists of three steps, each of which removed variability on decreasing time scales while preserving the flare signal:  

\begin{enumerate}
\item We fit and subtract a third order spline function that goes through the start and end of any light curve portion that has no gaps longer than 2 h, and through an averaged flux point every 6 h in between\footnote{30 h for stars less active than AU Mic}. This step removes long term trends as well as starspot variability on time scales of several days. If the light curve portion is shorter than 5 d, this step is skipped.
\item We iteratively remove strong periodic signal on time scales between 2 h and 5 d from the light curve. Each iteration first masks outlier points using a padded sigma-clipping procedure. For this step, single outliers above $3.5\sigma$ are masked as pure outliers. Series of $n>1$ data points above $3.5\sigma$ are masked as flare candidates, and padded with rounded $\sqrt{n}$ masked points before the outliers to capture slow rise phases, and rounded $2\sqrt{n}$ after the series to capture a potential extended decay phase that flares often display. Then we calculate a Lomb-Scargle periodogram~\citep{lomb1976,scargle1982} for the light curve, and perform a least-square fit with a cosine function using the dominant frequency in the periodogram as a starting point. The cosine fit is then subtracted from the light curve. We iterate five times or until the dominant peak's signal-to-noise ratio drops below 1.
\item Finally, we again apply the padded outlier clipping, and smooth any remaining variability that is not sinusoidal, first with a 6 h and then with a 3 h window 3rd order Savitzky-Golay~\citep{savitzky1964} filter implemented in \texttt{lightkurve} as \texttt{LightCurve.flatten}.
\end{enumerate}

These three steps can sometimes miss the very edges of the light curve, leaving small exponential drops or rises in the flux that affect the quiescent flux level calculation and/or produce false positive flare detections. If the first or the last data point is a $1\sigma$ outlier in the de-trended light curve, we fit an exponential growth or decay function to these fringes.

Finally we estimate the noise in the de-trended light curve using a rolling standard deviation with a 2 h window after padding outliers in the aforementioned way, but now above $1.5 \sigma$\footnote{$2.5 \sigma$ for stars less active than AU Mic}. We interpolated the masked outliers to arrive at a noise in the flare regions that is informed by the flux uncertainty in an adjacent light curve portion.

\begin{figure*}
\includegraphics[width=\hsize]{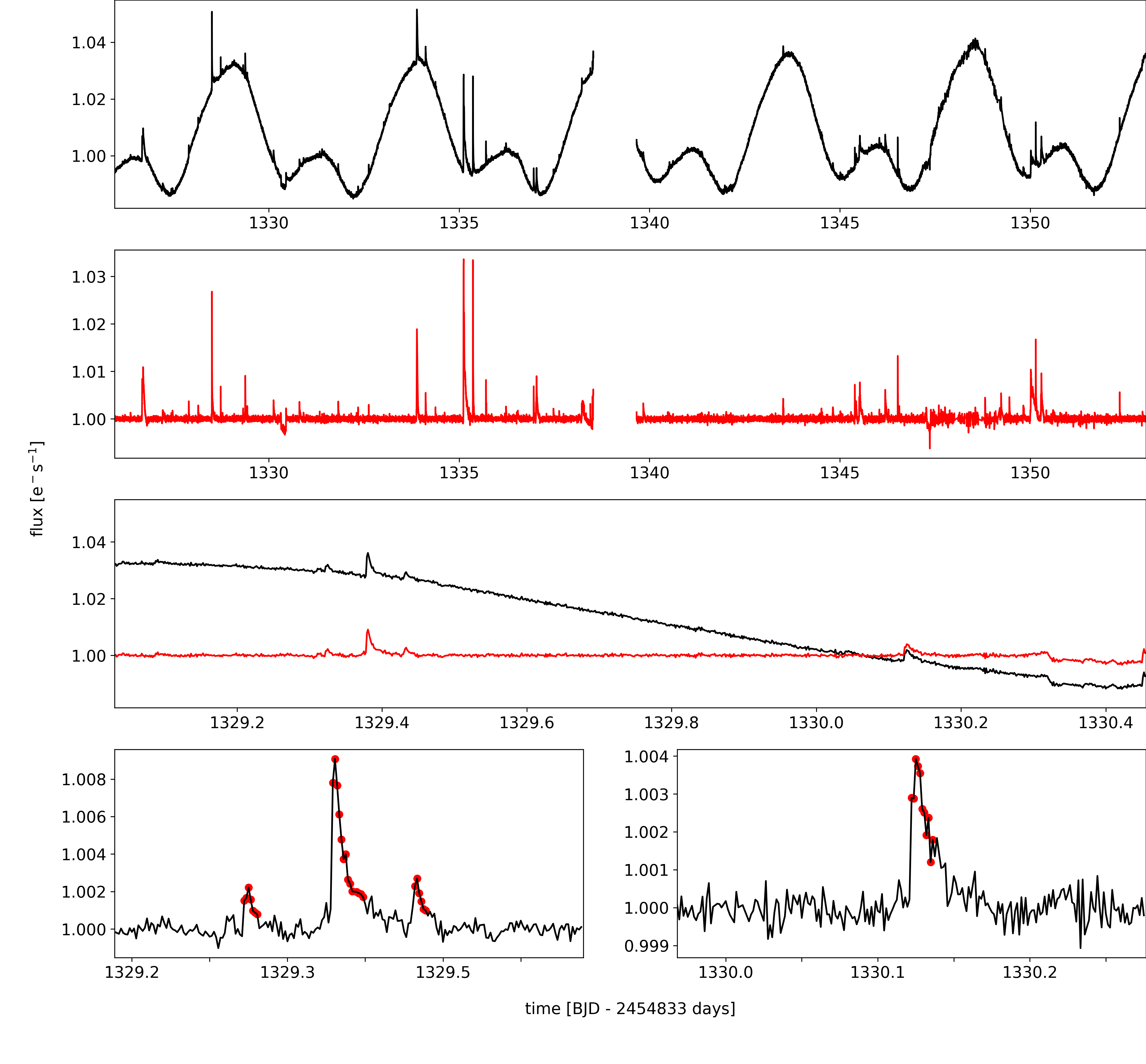} 
\caption{Subset of flares detected in the TESS light curve of AU Mic, Sector 1. Top panel: \texttt{PDCSAP\_FLUX} light curve. Second panel from the top: De-trended light curve. Third panel from the top: section from the \texttt{PDCSAP\_FLUX} (black) and de-trended (red) light curve. Two bottom panels: de-trended light curves of flares confirmed in the panel above. The red dots indicate the data points that mark the flares.}
\label{fig:illustrate_detrend}
\end{figure*}

The result for the light curve in Sector 1 is shown in Fig.~\ref{fig:illustrate_detrend}. This method is based on \texttt{AltaiPony}~\citep{ilin2021altaipony}, an open-source Python toolkit for flare-focused light curve analysis of Kepler and TESS data, and was inspired by the iterative approach in~\citet{davenport2016}, who searched the entire Kepler catalog for flares. We tested this method on a variety of synthetic and real light curves in the TESS and Kepler archives, so that it can be applied to a larger sample. We provide the de-trending module and an example script for the use of this method with TESS short cadence light curves in this project's Github repository\footnote{\url{https://github.com/ekaterinailin/flaring-spi/tree/master/notebooks}}.

\subsubsection{Flare finding}
\label{sec:flarefind}
We searched the de-trended light curves for flare candidates. For each light curve, we first find an iterative median, and then apply the threshold method introduced by~\citep{chang2015} that requires three consecutive positive outliers $3 \sigma$ above median for a candidate detection. To these data points we then add subsequent data points until one of them falls below a $2 \sigma$ above median threshold to avoid cutting off detectable parts of the flare decay phase. This series of data points is then flagged as a flare candidate. For each flare, the pipeline returns the flare start and end points, duration, amplitude, and equivalent duration ($ED$) with uncertainties. The $ED$ is the integrated flare flux $F_{flare}$ divided by the median quiescent flux $F_0$ of the star, integrated over the flare duration~\citep{gershberg1972}:
\begin{equation}
\label{eq:ED}
ED=\displaystyle \int \mathrm dt\, \frac{F_{flare}(t)}{F_0}.
\end{equation}

We tested our flare-finding procedure on a range of both real and synthetic light curves that covers several typically observed spot-induced variability signal patterns, and that contains flare signatures between ones that barely exceed the detection threshold to the largest flares we typically observe. Since the light curve of AU Mic shows variability not only with rotation but also on shorter time scales comparable to those of flares, we confirmed all flare candidates by eye. In cases where the flare shape was not well captured by the algorithm, we manually corrected the flare duration by adding or subtracting data points to or from the detection. We treated the transit light curves of AU Mic b and c manually, too. Spot occultations by the transiting planet and small flares have similar shapes and time scales, so that we picked flares that occurred within transits directly by eye on a case by case basis. 



\subsubsection{Flare catalog}
\label{sec:flarecatalog}
\begin{table}
\caption{Confirmed flare events in the TESS light curves of AU Mic, sorted by orbital phase of AU Mic b. The remainder of the table is available in electronic form.}
\centering
\begin{tabular}{l|ccccc}
\hline
 Sec. &  $t_s$ [BTJD] &  $t_f$ [BTJD] &  orb. phase &    $a$ &         $ED$ [s] \\
\hline
   27 &     2058.2378 &     2058.2526 &       0.001 &  0.007 &  $3.48 \pm 0.04$ \\
   27 &     2058.2584 &     2058.2593 &       0.004 &  0.002 &  $0.10 \pm 0.01$ \\
   27 &     2041.3551 &     2041.3572 &       0.006 &  0.002 &  $0.23 \pm 0.02$ \\
    1 &     1330.4514 &     1330.4709 &       0.007 &  0.003 &  $1.65 \pm 0.03$ \\
   27 &     2041.3690 &     2041.3734 &       0.008 &  0.003 &  $0.70 \pm 0.03$ \\
   27 &     2041.6186 &     2041.6202 &       0.038 &  0.002 &  $0.15 \pm 0.02$ \\
   27 &     2050.1252 &     2050.1259 &       0.043 &  0.003 &  $0.10 \pm 0.01$ \\
   27 &     2050.1724 &     2050.1731 &       0.048 &  0.001 &  $0.07 \pm 0.01$ \\
    1 &     1330.8028 &     1330.8250 &       0.049 &  0.002 &  $1.72 \pm 0.06$ \\
   27 &     2058.7093 &     2058.7148 &       0.057 &  0.004 &  $0.74 \pm 0.03$ \\
\hline

\end{tabular}

\label{tab:flares}
\end{table}

\begin{figure}
\includegraphics[width=\hsize]{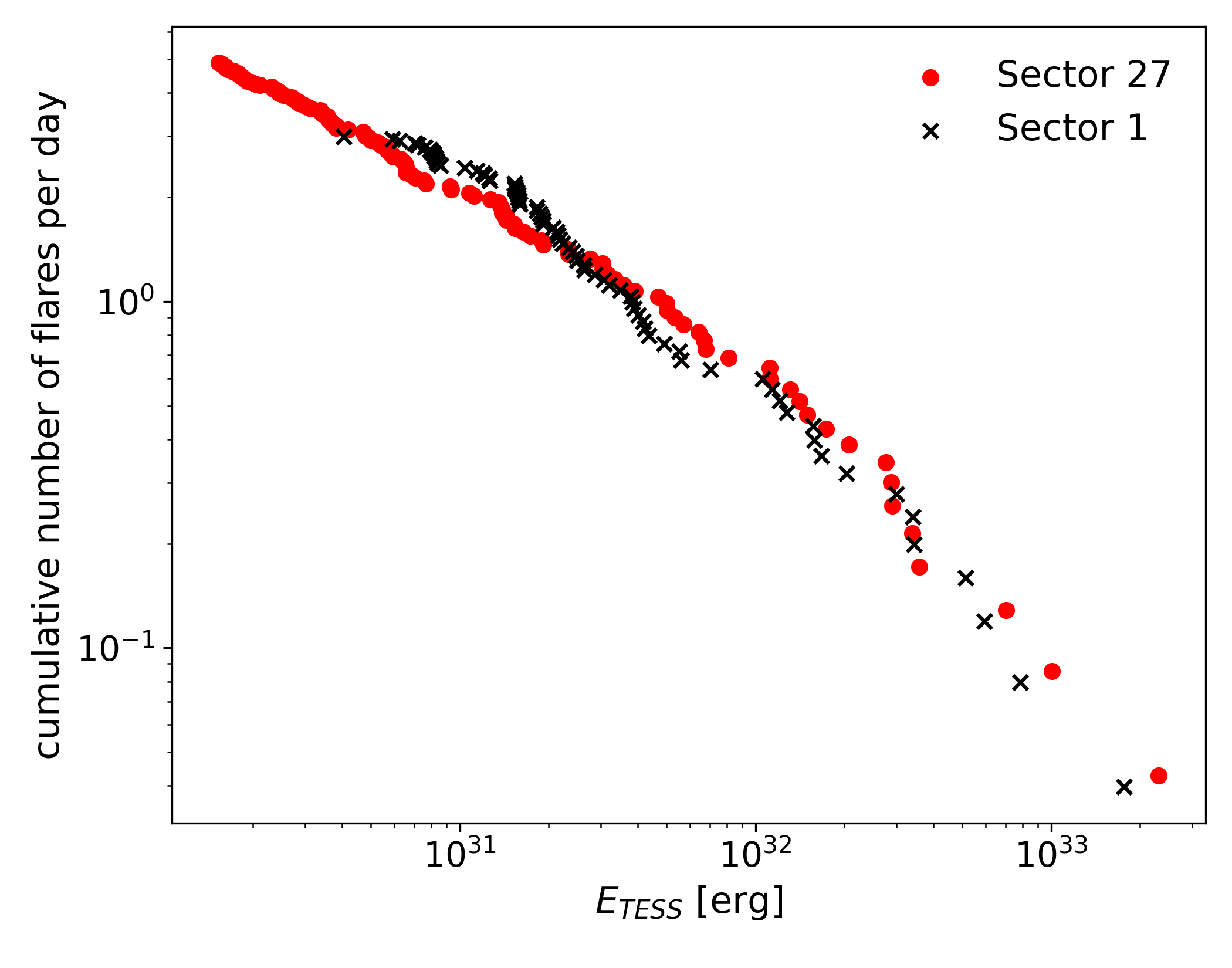} 
\caption{Cumulative flare frequency distributions (FFDs) obtained from the two TESS light curves of AU Mic. $E_{TESS}$ is the flare energy emitted in the TESS band.}
\label{fig:ffd}
\end{figure}
We confirmed 75 flares in Sector 1, and 114 flares in Sector 27~(see bottom panels in Fig.~\ref{fig:illustrate_detrend} for examples). We found more flares in Sector 27 because its six times higher observing cadence lowers the detection threshold, which can be seen in the cumulative flare frequency distributions for both Sectors~(Fig.~\ref{fig:ffd}). Overall, we confirmed fewer flares than reported by \citet{martioli2021new} who found 162 and 157 flares in Sector 1 and 27, respectively. However, we confirmed almost twice as many flares as \citet{gilbert2021flares}. We attribute these differences to the respective detection methods. \citet{martioli2021new} flagged single $2.5\sigma$ outliers as candidates, while we required three consecutive data points $3\sigma$ above the noise. \citet{gilbert2021flares} used a Bayesian template comparison method to detect probable candidates, which slides a Gaussian rise and exponential decay flare template over the data, and returns the probability of belonging to a flare for each data point. Therefore, their method selects flares that fit the template even if they are small, while our method is more senstive to flares that do not conform to the classical shape but misses flares that don't pass the sigma threshold. Table \ref{tab:flares} lists the confirmed flare events with their respective start and finish times $t_s$ and $t_f$, orbital phase of AU Mic b at $t_s$, amplitude $a$ and $ED$.

\section{Flare rate variation with rotational, orbital, and synodic phases}
\label{sec:phases}

\begin{figure}
\includegraphics[width=\hsize]{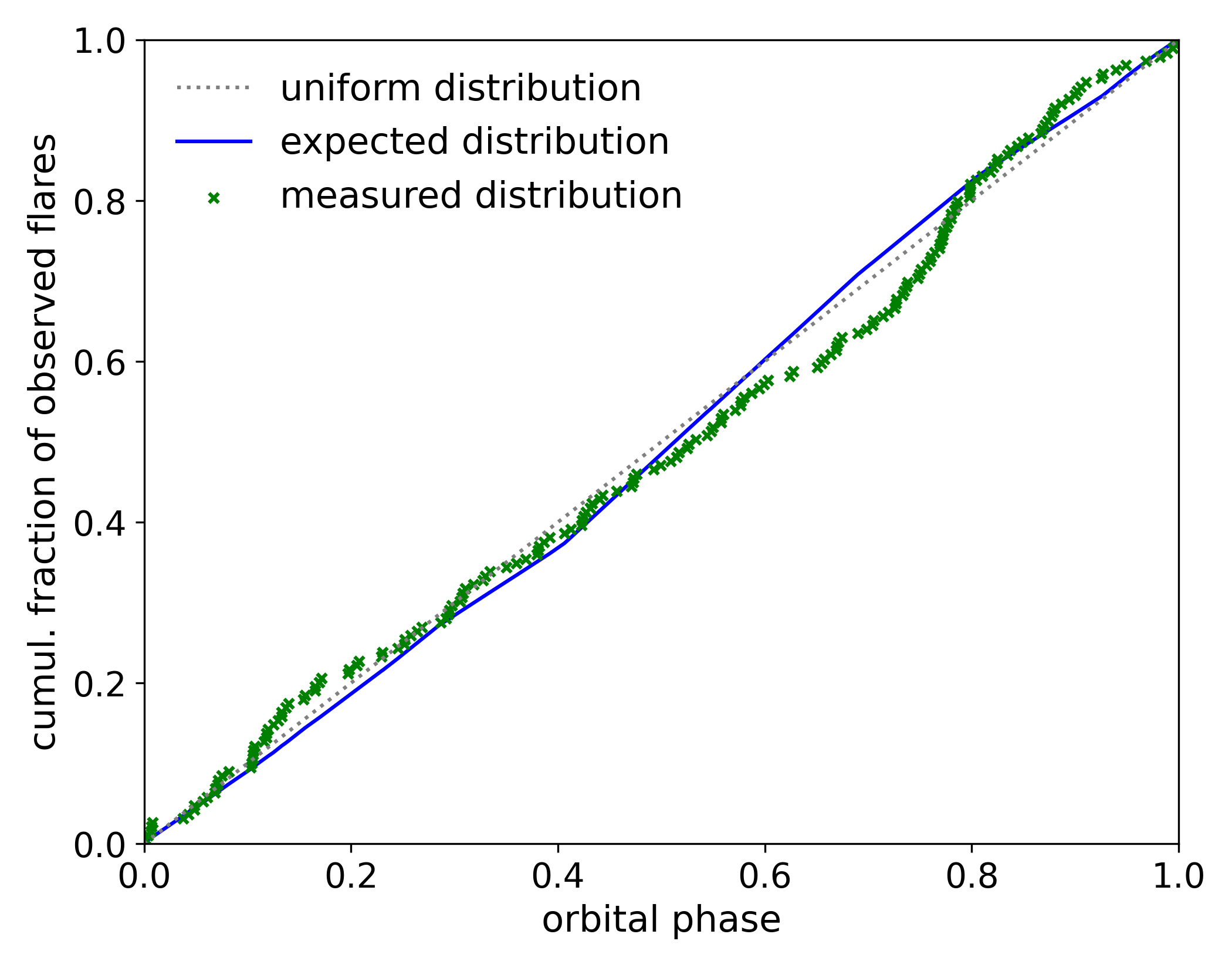} 
\caption{Cumulative distribution of the full sample of flares with orbital phase of AU Mic b (Sectors 1 and 27, green crosses). Phase 0 indicates the planet's transit mid-time. The underlying distribution is uniform (grey dotted line). In the data, however, it is modulated by gaps in the phase coverage and the different detection thresholds of the two light curves. The modulated distribution (blue line) is used as null hypothesis in the A-D test.}
\label{fig:cumdist}
\end{figure}

\begin{table}
\caption{Median $p$-values of the custom A-D tests for the orbital, rotational and synodic periods of AU Mic (and AU Mic b) calculated using 20 different start phases. Smallest $p$-value is boldfaced. There is no significant deviation from uniform flaring in time with either of the periods. $n$: number of flares in sample.}
\centering
\begin{tabular}{llcccc}
\hline
          &  &      & $p(P_{orb})$ & $p(P_{rot})$ & $p(P_{syn})$ \\
Sec. & sample &     $n$ &              &              &              \\
\hline
both & $ED>1\,$s &   71 &       $\mathbf{0.07}$ &       $0.79$ &       $0.37$ \\
          & $ED<1\,$s &  118 &       $0.52$ &       $0.78$ &       $0.75$ \\
          & total &  189 &       $0.21$ &       $0.57$ &       $0.43$ \\
1 & $ED>1\,$s &   38 &       $0.27$ &       $0.71$ &       $0.48$ \\
          & $ED<1\,$s &   37 &       $0.56$ &       $0.63$ &       $0.44$ \\
          & total &   75 &       $0.20$ &       $0.64$ &       $0.71$ \\
27 & $ED>1\,$s &   33 &       $0.44$ &       $0.11$ &       $0.48$ \\
          & $ED<1\,$s &   81 &       $0.52$ &       $0.53$ &       $0.68$ \\
          & total &  114 &       $0.53$ &       $0.24$ &       $0.68$ \\
\hline

\end{tabular}

\label{tab:pvals}
\end{table}

In analogy to solar flares, stellar flares are thought to occur in the vicinity of active regions on the stellar surface, where magnetic field lines emerge, and magnetic energy can accumulate. AU Mic is a young, extremely active flaring star that is covered with large active regions~\citep{linsky1994, kochukhov2020, plavchan2020}, and possibly a multitude of magnetic loops across the entire corona~\citep{cranmer2013}.

Flares on early M dwarfs tend to occur randomly in the various active regions which cover the stellar surface~\citep{doyle2018, doyle2019}. A variation of flaring rate with stellar rotational phase can be caused by an extremely active region that rotates in and out of view of the observer. A very inactive region that suppresses flaring compared to the rest of the stellar surface is also conceivable, for instance in a region wherein the magnetic field is strong enough to sufficiently suppress convection that is required to produce flares. In both cases, flare rates would be elevated or decreased for a fixed fraction of the rotation period, which depends on the inclination of the star and the latitude of the active region. This signal is then smeared out by differential rotation~\citep{howard2021evryflare}. Notably, a non-uniform rotational phase distribution of flares involves only the star. 

When a magnetized planet that orbits inside the Alfv\'en surface of the star comes into play, it can introduce a deviation from a uniform flare distribution in time via magnetic star-planet interaction. The magnetic field lines that connect the star's to the planet's magnetic field channel energy from the planet to the star, that is then dissipated via flares. The footpoint of the connection field lines move across the stellar surface as the planet orbits the star. If the stellar properties at the location of these footpoints are negligible or sufficiently uniform not to affect the interaction, excess flares can be observed whenever the footpoint is in view, in phase with the orbital period $P_{orb}$. If, however, the stellar properties at the footpoint's location matter, this occurrence of excess flares will be modulated with the stellar rotation period $P_{rot}$. Then the relevant period is the synodic period $P_{syn}$~\citep{fischer2019}, with

\begin{equation}
P_{syn} = \left|\dfrac{1}{P_{rot}} - \dfrac{1}{P_{orb}}\right|^{-1}.
\end{equation}

In summary, if an inhomogeneous distribution of active regions on the stellar surface, or the interaction of AU Mic with the innermost planet takes place, triggering flares in the observable regime, the flaring rates in phase with $P_{rot}$, $P_{orb}$ or $P_{syn}$  will deviate from a uniform distribution. Our null hypothesis is that no rotational, orbital, or synodic dependence is present. We test if we have to reject this hypothesis using a customized Anderson-Darling test~\citep[\mbox{A-D} test,][]{anderson1952, stephens1974edf}, a non-parametric goodness-of-fit test suited for continuous variables, such a flare frequencies. The \mbox{A-D} test is more sensitive to the ends of the distribution, i.e. phases near 0 or 1, than the more widely used Kolmogorov-Smirnov~\citep{kolmogorov1933sulla,smirnov1948table} test. Since the starting phase should ideally not affect the outcome of the test, we choose the \mbox{A-D} test.

With this method we can analyse $N$ light curves that cover observing periods that do not overlap (to avoid double counting of flares), and each of which has the same flare detection threshold throughout the observing time of the light curve. It is permissible, however, that this threshold varies from light curve to light curve, as is the case in our data set for AU Mic.
 
After having searched all light curves for flares~(see Section~\ref{sec:flarecatalog}), we calculate the deviation from a uniform underlying flare rate distribution with orbital, rotational, and synodic phase each using a customized A-D test in six steps:
\begin{enumerate}
\item Calculate $\phi$, the expected frequency distrbution of flares per phase (expected distribution for short), using all valid data points in the de-trended light curves, taking into account the different observing cadences, and detection thresholds.
\item Sample $N$ flare occurrence phases $x_i$ with $i \in [1,N]$ from the expected distribution $\phi$ to get a distribution of flare phases, where $N$ is the total number of flares in the data. This step generates a list of flares that are uniformly distributed across all phases, modulated only by the phase coverage of the observations, and the detection thresholds of the light curves. This list is a sample from the expected distribution we test against~(see Fig.~\ref{fig:cumdist}).
\item Repeat the previous step to obtain a large number of samples, and calculate the A-D statistic $A^2$ for each of them using the cumulative form $\hat{\phi}(x_i)$ of the expected distribution, where the $x_i$ are sorted in ascending order. 

\begin{equation}
A^2 = - N -\displaystyle\sum_{i=1}^N \dfrac{2i-1}{N}\left(\log(\hat{\phi}(x_i))+\log(1-\hat{\phi}(x_{N+1-i}))\right)
\end{equation}

In this study, we generated $10\,000$ samples for each test. We also varied the start phase of the distribution between 0 and 1 with a step size of 0.05. While the A-D test is less sensitive to the start phase than the K-S test, we still observed a scatter in the outcomes of the tests. We accounted for this by calculating the median of all start phase varied test results.
\item Take all $N$ flare occurrence times from the TESS data, and derive their phases $x_{i,obs}$ with $i \in [1,N]$. 
\item Sort all $x_{i,obs}$ by phase in ascending order using the same start phase as the expected distribution.
\item Finally, compare the $A^2_{obs}$ value of the observed distribution of flare phases against the expected distribution of $A^2$, and calculate the significance level ($p$-value) of the difference. 
\end{enumerate}

The $p$-values for the A-D test obtained from the above procedure are shown in Table~\ref{tab:pvals}. No choice of Sector, energy split, or period resulted in a consistently significant deviation from uniform flaring in time. The strongest deviation from uniformity, a $\sim1.5\sigma$ detection ($p=0.07,\;n=71$), is seen with $P_{orb}$ in the flares whose energy exceeds $ED=1$\,s, that is, approximately $2.3\cdot10^{31}\,$erg in the TESS band. The $p$-values in this subsample were consistently $<0.175$ at all 20 start phases, which is not the case for any other subsample in Table~\ref{tab:pvals}.

The observed distribution of flares does not deviate significantly from a uniform distribution of flares with orbital or synodic period. \citet{howard2021evryflare} detected flaring periodicity in 3 out of 284 late K and M dwarfs observed with TESS, and concluded that it may be rare, difficult to detect, or both.

Because AU Mic is seen nearly equator-on, the non-detection with $P_{rot}$ implies a uniform distribution of flares with stellar longitude. The spot structure evolved between Sector 1 and 27~\citep{martioli2021new}, but during each Sector spot patterns were stable over several rotation periods of the star~\citep{szabo2021changing}. However, considering each Sector separately does not reveal a significant deviation from uniform flaring with longitude. This observation is consistent with most M dwarfs that were searched for rotational periodicity~\citep{doyle2018, doyle2019}.

We also applied the above method to test for active longitudes excited by the close 4:7 $P_{rot}$:$P_{orb}$ commensurability~\citep{szabo2021changing,cale2021diving} of AU Mic and AU Mic b. The excitation of resonant oscillations in the stellar magnetic field of close-in star-planet systems was predicted by~\citet{lanza2022model}. We found no significant deviation from uniformity with the resonance period $P_{rot}/4$. However, for the AU Mic system specifically,~\citet{lanza2022model} estimate that this excitation will not occur due to the planet's relatively large distance from its host, consistent with our result.
 

\section{Discussion}
\label{sec:discussion}
\begin{figure}
\includegraphics[width=\hsize]{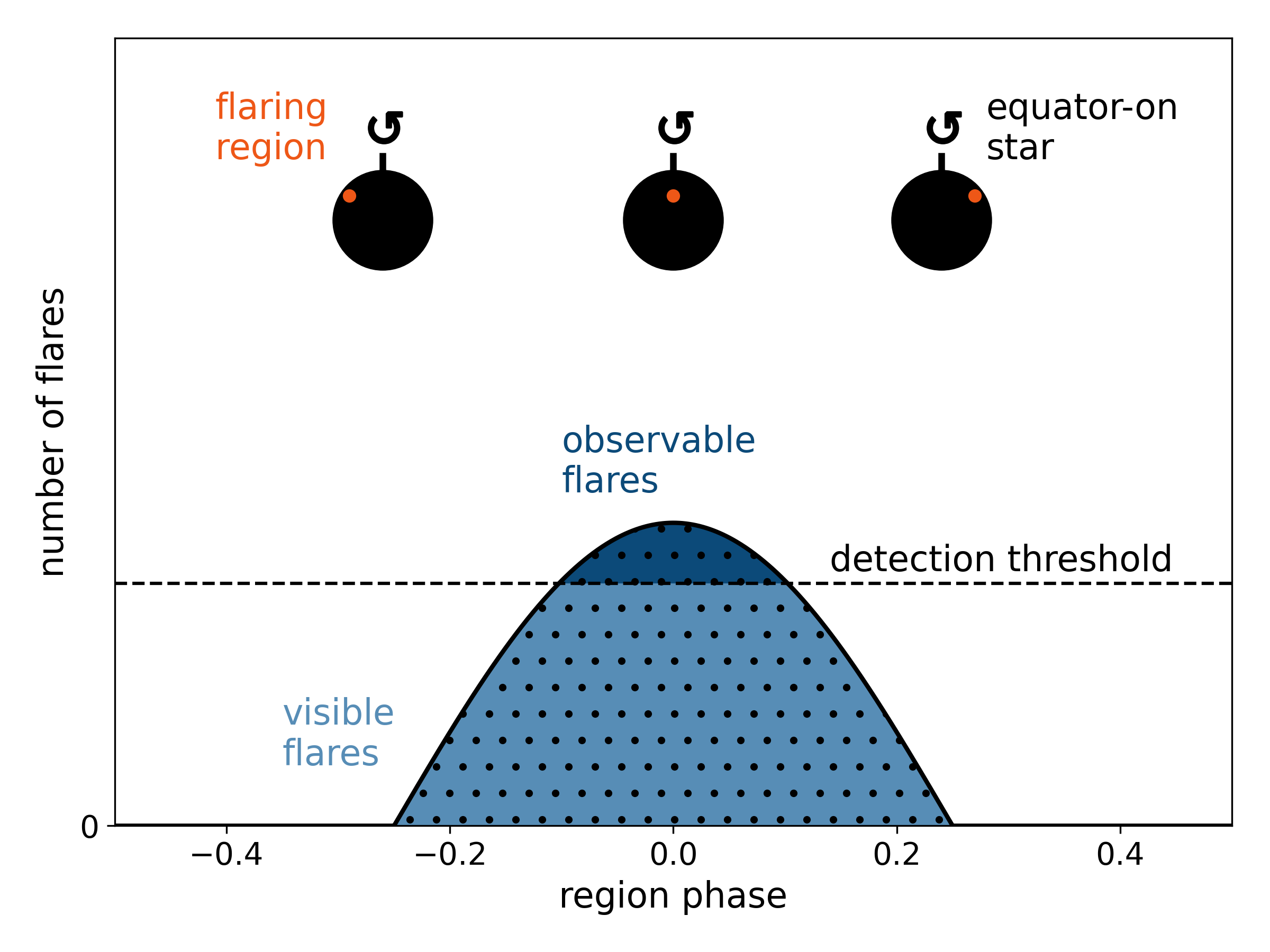} 
\caption{Number of flares produced by a single flaring region on the stellar surface at different phases of the region. The flares become visible (light blue area) when the region comes into view. However, they are not observable because the amplitude is geometrically foreshortened when the region is close to the limb. When the region approaches the center of the stellar disk, the flares' amplitude exceeds the detection threshold so that they can be observed (dark blue area). The phase can be the stellar rotation phase, the orbital phase of the planet, or the synodic phase of the star-planet system.}
\label{fig:observability}
\end{figure}

If the tentative signal with orbital phase \mbox{($p=0.07,\;n=71$)} truly stems from flaring SPI with AU Mic b, we can estimate the required observing time to confirm the detection. In our estimate, we doubled and tripled the sample by introducing one and two additional flares between each pair $(x_i,x_{i+1}),\;i \in [1,N]$, of real flares, placing them at equal distance from each other and smearing out the exact occurrence phase with a Gaussian with a standard deviation of $(x_{i+1}-x_i)/2$ and $(x_{i+1}-x_i)/3$, respectively. We executed the A-D test procedure from Section~\ref{sec:phases} on the new $2N$ sized flare distribution, sampling $2\cdot 10^5$ and $3\cdot 10^5$ $A^2$ values for each start phase, instead of $10^4$, to obtain the critical values in the high significance tail of the $A^2$ distribution. Assuming that the structure of the phase distribution remains the same over time, doubling the sample size yields $p\approx 0.003$; tripling yields $p\approx0.0008$. Therefore, since the deviation was found in the higher energy flares in our sample, monitoring AU Mic for another $50-100\,$d with instruments with the same or somewhat lower photometeric precision and observing cadence as TESS may suffice to detect flaring SPI in this system with $>3\sigma$ significance.

If the signal is a fluke in the data, the observed absence of flaring SPI signal can be explained in two ways -- observational biases that impede detection, and physical reasons that prevent flaring SPI to occur in the first place. We consider three physical explanations first.

A first physical reason could be the unknown magnetic field of AU Mic b. It is possible that AU Mic b has no magnetic field that can interact with the stellar magnetic field to trigger flares~\citep{lanza2018close-by}. A magnetic dynamo driven by convective motion of conductive fluids in the interior of the rotating planet is the most efficient mechanism to produce the magnetic field strengths observed in Solar System planets. At the same time, planetary magnetic fields in the Solar System are diverse, which is thought to depend heavily on the internal structure and composition of the planet~\citep{stevenson2003planetary, jones2011planetary}. Only a few indirect measurements of exoplanet magnetic fields exist to date~\citep[e.g.][]{cauley2019magnetic, ben-jaffel2021signatures}. Therefore, it is difficult to predict the presence or the strength of AU Mic b's field based solely on mass and radius estimates.

Second, AU Mic b's orbit could be outside the Alfv\'en zone of AU Mic, either constantly or temporarily. Although the star has a strong magnetic field, a high mass loss rate may pull the Alfv\'en surface closer to the star~\citep{kavanagh2021}, so that no energy can be channelled from the planet to the star. Following \citet{kavanagh2021}, the absence of SPI would suggest that the mass loss rate of AU Mic was closer to $1000\dot{M}_\odot$ than to $10\dot{M}_\odot$. This situation may change with evolving stellar cycle.~\citet{ibanezbustos2019first} report a 5-year chromospheric activity cycle in AU Mic, but it is not clear how it affects mass loss.

Third, instead of via flares, energy may be transferred between planet and star gradually.~\citet{shkolnik2003evidence, shkolnik2005hot}, and~\citet{shkolnik2008nature} observed evidence of chromospheric hot spots in several systems with close Hot Jupiter companions that could partially be explained in terms of magnetic SPI~\citep{cohen2011dynamics,lanza2012starplanet}. It is not clear if flares and hot spots could be effects of the same SPI reconnection event or instead represent separate channels. The chromospheric observations in Shkolnik et al. did not provide the observing cadence required to conclusively detect or exclude coincident flares in excess of intrinsic flaring activity.

Observational biases can also prevent the detection of flaring SPI. We discuss four possible biases here, all of which can be interpreted within Fig.~\ref{fig:observability}:

First, the effect might simply be too weak. SPI flares may be triggered rarely, with only a handful of flares in our sample, which are not detectable with sufficient significance. This corresponds to fewer dots per area, i.e. flares per time, in Fig.~\ref{fig:observability}.

Second, we may be looking into the wrong flare energy regime. Flaring SPI may manifest at low energies $<10^{30}$ erg, i.e. below the detection threshold (no dark blue area in Fig.~\ref{fig:observability}). If, on the contrary, SPI flares are triggered at very high energies, they may, again, occur too rarely to be observed within the given monitoring time. 

Third, flaring activity could be elevated through flaring SPI, but manifest as excess flares across all longitudes. While the mechanism of flaring SPI is not well understood, we can imagine circumstances that create a large or spread-out SPI-induced flaring region, which would always be in view. Flaring region sizes increase with flare energy for intrinsic flares~\citep{sammis2000dependence, notsu2019kepler, howard2020evryflare}. It is therefore conceivable that extreme amounts of released energy during SPI might manifest in large flaring regions. Another scenario is efficient transport of the released energy or the magnetic field perturbation within the stellar magnetosphere before erupting in a flare. This could result in a broad range of longitudes at which SPI flares occur. Alternatively, the magnetic field lines connecting planet and star could be arranged such that the locations they map to on the stellar surface are distributed in a broad longitudinal range.

Finally, flares triggered by the interaction might be frequent enough and have sufficiently high amplitudes to be detected if the flaring region was located close to the equator. Yet if the region was at high latitudes, flares would always appear geometrically foreshortened, and remain below the detection threshold.

However, if the tentative $1.5\sigma$ signal in the $ED>1$\,s flares~(Table~\ref{tab:pvals}) indeed stems from flaring SPI, we may be limited mostly by the first observational bias, i.e., insufficient observing time. In this case, the flares triggered by the interaction are well above the detection threshold of TESS.

\section{Conclusions}
\label{sec:conclusions}
Since SPI flares are expected to be morphologically identical to intrinsic stellar flares, a statistically significant orbital phase dependence of flaring behaviour is the closest we can currently get to detecting the effect directly. Using high cadence TESS observations of AU Mic, an early M dwarf with a close-in planet detected in 2020, we did not find any sign of flaring SPI.

The strength of the interaction is difficult to constrain in theory~\citep{strugarek2019}, and multiple physical models of flaring SPI co-exist~\citep{lanza2018close-by, saur2013magnetic}. Moreover, the orbit of AU Mic b is not yet synchronized with the stellar rotation period which may additionally lead to tidal interaction~\citep{cuntz2000stellar}. Although AU Mic shows the largest number of flares detected among all exoplanet hosts found by Kepler and TESS to date, we conclude that most of them are manifestations of intrinsic stellar activity. 

We do not rule out that, as time series observations of AU Mic accumulate with TESS, and soon PLATO~\citep{rauer2014plato}, SPI-triggered flares can be detected with high significance. We found the strongest deviation \mbox{($p=0.07,\;n=71$)} from intrinsic flaring with the orbital period $P_{orb}$ of AU Mic b, in the high energy half of our flare sample ($ED>1$\,s). We estimate that extending the observing time by a factor of $2-3$ will yield a $>3\sigma$ detection, if the temporal structure of the signal is preserved over time. We argue that long gaps on time scales of months to years betweeen consecutive light curves will average out rotational variability signal as the spot structure evolves, but will leave a $P_{orb}$-dependent SPI signal intact. In contrast, a persisting absence of flaring SPI would support models that favor an Alfv\'en zone inside AU Mic b's orbit; strong winds; weak planetary magnetic fields; or a more gradual energy transfer between planet and host.

In an upcoming study, we will apply the methods of flare finding and orbital phase analysis presented here to a large sample of star-planet systems with known close-in planets. This will increase the total number of orbits covered, and probe the effects of orbital distance and star-planet mass ratio on the presence and intensity of flaring SPI.
\section*{Acknowledgements}
The authors would like to thank Gyula M. Szab\'o for comments that improved the quality of this work. 
EI would like to thank Florian U. Jehn for advice on data visualization. EI acknowledges support from the German National Scholarship Foundation. KP acknowledges support from the German Leibniz Community under grant P67/2018. We made use of the Python packages numpy~\citep{numpy2020} and pandas~\citep{mckinney2010data,pandas2020software}, as well as GNU Parallel~\citep{tange2018gnu}. This project made use of computational systems and network services at the American Museum of Natural History supported by the National Science Foundation via Campus Cyberinfrastructure Grant Award \#1827153 (CC* Networking Infrastructure: High Performance Research Data Infrastructure at the American Museum of Natural History). This research has made use of the SIMBAD database, operated at CDS, Strasbourg, France~\citep{wenger2000}. This paper includes data collected with the TESS mission, obtained from the MAST data archive at the Space Telescope Science Institute (STScI). Funding for the TESS mission is provided by the NASA Explorer Program. STScI is operated by the Association of Universities for Research in Astronomy, Inc., under NASA contract NAS 5-26555.

\section*{Data Availability}
TESS light curves are publicly available through the Mikulski Archive for Space Telescopes (\url{https://mast.stsci.edu/portal/Mashup/Clients/Mast/Portal.html}). The python scripts and modules used to permorm the statistical analysis and generate the figures in this work as well as the full versions of Tables \ref{tab:flares} and \ref{tab:pvals} can be found on Github~(\url{https://github.com/ekaterinailin/flaring-spi}).




\bibliographystyle{mnras}
\bibliography{bibliography}




%
%


\bsp	
\label{lastpage}
\end{document}